\documentclass{aastex701}

\usepackage{makecell}

\begin{document}

\title{Unbound Tails and Compressed Heads: A JCMT Study of the SFO 38 Cloud}

\author[orcid=0009-0002-6147-531X]{Puja Porel}
\affiliation{Indian Institute of Astrophysics, II Block, Koramangala, Bengaluru 560034, India}
\affiliation{Pondicherry University, R.V. Nagar, Kalapet, 605014, Puducherry, India}
\email[show]{pujaporel11@gmail.com} 

\author[orcid=0000-0002-6386-2906]{Archana Soam} 
\affiliation{Indian Institute of Astrophysics, II Block, Koramangala, Bengaluru 560034, India}
\email{archana.soam@iiap.res.in}

\author[orcid=0000-0001-5996-3600]{Janik Karoly}
\affiliation{Jeremiah Horrocks Institute, University of Central Lancashire, Preston PR1 2HE, UK}
\email{JKaroly@uclan.ac.uk}

\author[orcid=0000-0003-0014-1527]{Eun Jung Chung}
\affiliation{Korea Astronomy and Space Science Institute (KASI), 776 Daedeokdae-ro, Yuseong-gu, Daejeon 34055, Republic of Korea}
\email{rigelej@gmail.com}

\author[orcid=0000-0002-3179-6334]{Chang Won Lee}
\affiliation{Korea Astronomy and Space Science Institute (KASI), 776 Daedeokdae-ro, Yuseong-gu, Daejeon 34055, Republic of Korea}
\affiliation{University of Science and Technology, Korea (UST), 217 Gajeong-ro, Yuseong-gu, Daejeon 34113, Republic of Korea}
\email{cwl@kasi.re.kr}

\author[orcid=0000-0001-9333-5608]{Shinyoung Kim}
\affiliation{Korea Astronomy and Space Science Institute (KASI), 776 Daedeokdae-ro, Yuseong-gu, Daejeon 34055, Republic of Korea}
\email{shinykim@kasi.re.kr}

\author[orcid=0000-0002-8614-0025]{Shivani Gupta}
\affiliation{Indian Institute of Astrophysics, II Block, Koramangala, Bengaluru 560034, India}
\affiliation{Pondicherry University, R.V. Nagar, Kalapet, 605014, Puducherry, India}
\email{shivani.gupta@iiap.res.in}

\author{Neha Sharma}
\affiliation{Aryabhatta Research Institute of Observational Sciences (ARIES), Nainital 263001, India}
\email{pathakneha.sharma@gmail.com}

\begin{abstract}

SFO\,38, located in the Cepheus molecular cloud within the northern part of the H\,\textsc{ii} region IC\,1396, is shaped by intense ultraviolet radiation from the nearby O6.5V-type star HD\,206267 and represents a classic example of a bright-rimmed cloud (BRC) undergoing radiatively driven implosion (RDI). While previous studies have examined the southern globule using CS and $^{13}$CO (1--0), we present a refined analysis using high-resolution JCMT-HARP observations in the $^{12}$CO, $^{13}$CO and C$^{18}$O ($J = 3 \rightarrow 2$) lines, deriving key physical parameters along with virial mass and turbulence properties of the southern head. We also perform the first detailed investigation of the northeastern and northwestern tails, determining their morphological dimensions and internal conditions, including excitation temperature, column density, mass, and volume density. Spectral and stability analyses reveal that the tail regions are gravitationally unbound and dynamically expanding, explaining the lack of active star formation. Our results further shed light on the evolutionary fate of these structures, assessing whether they may accumulate sufficient material to become future sites of star formation or remain quiescent. Overall, this work highlights the dual role of RDI in this BRC: while it triggers star formation in the dense head, it simultaneously disperses and dynamically reshapes the extended tails.

\end{abstract}



\section{Introduction}
\label{sec:intro}

\begin{figure*}
\begin{center}
\resizebox{18.0cm}{18.0cm}{\includegraphics{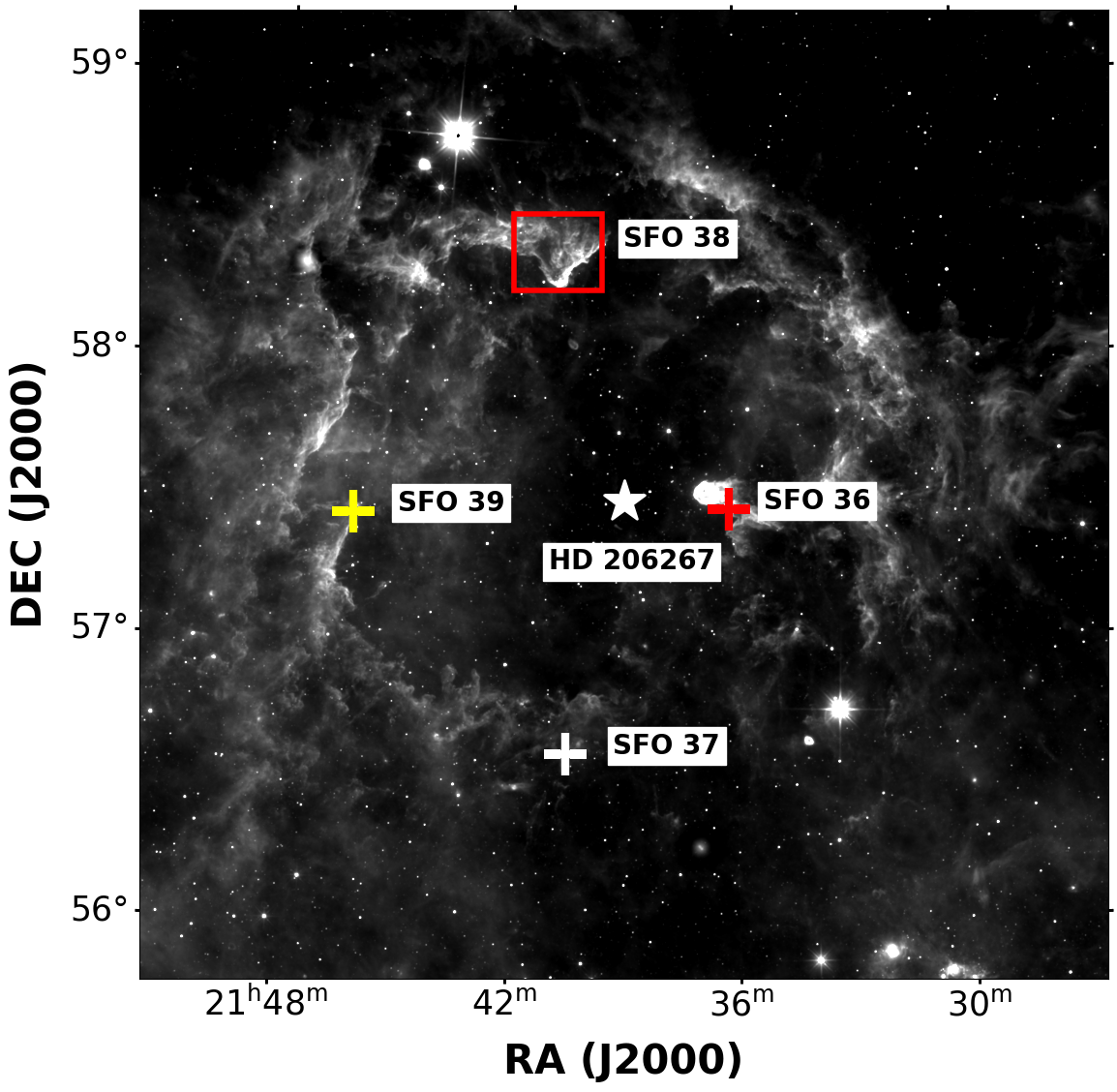}}
\caption{The WISE 12-micron intensity map, encompassing a region approximately 3.4$^\circ \times$ 3.4$^\circ$ in a two-dimensional projection, is displayed at an angular resolution of 6.5$\arcsec$. The map highlights the locations of prominent bright-rimmed clouds: SFO 36, SFO 37, and SFO 39, marked by red, white, and yellow plus symbols, respectively. A red rectangular box, approximately 19$^\prime \times$ 16$^\prime$ in size, delineates the bright-rimmed cloud SFO 38, which has been observed with the JCMT telescope for $^{12}$CO, $^{13}$CO, and C$^{18}$O J = 3-2 transitions. The position of the ionizing source, the O6.5V-type star HD 206267, is marked with a white star symbol, signifying its pivotal role in driving the photoionization and shaping the morphology of SFO 38.}\label{Fig: wise and optical image}
\end{center}
\end{figure*}

Massive stars ($\geq 8 M_\odot$; \citealt{bally2008overview}) play a pivotal role in the process of star formation, not only through their own formation from the collapse of dense molecular clouds but also by significantly influencing their environment via powerful feedback mechanisms. The intense ultraviolet (UV) radiation, robust stellar winds, and eventual supernova explosions from these stars release vast amounts of energy into the surrounding interstellar medium. This feedback can trigger subsequent generations of star formation through various processes.

Two key models that describe how feedback from massive stars influences star formation in their vicinity are the \textit{Collect and Collapse} (C \& C) model \citep{elmegreen1977sequential, hosokawa2006dynamical, deharveng2003triggered, deharveng2009star} and the \textit{Radiation-Driven Implosion} (RDI) model \citep{ikeda2008sequential, figueira2020apex, saha2022investigation}. While both frameworks involve the interaction between massive stars and their surrounding gas, they differ in how this interaction induces gas collapse and subsequent star formation.

In the C \& C model, the formation of new stars is driven by the intense UV radiation from a massive star, which creates an expanding H\,\textsc{ii}  region. As this region grows, it sweeps up neutral gas, forming a dense shell surrounding the ionized region. The expansion of the H\,\textsc{ii} region compresses the surrounding molecular gas, increasing its density. Eventually, gravitational instabilities, such as the Jeans instability, occur within the shell, leading to the collapse of certain regions and the formation of dense molecular cores. This process triggers the birth of new stars. In this model, the feedback from the massive star compresses the surrounding gas, initiating sequential star formation, where each subsequent generation of stars is triggered by the expanding H\,\textsc{ii} region and its impact on the surrounding medium.

In contrast, the RDI model emphasizes the role of radiation pressure in driving the implosion of a molecular cloud. The intense UV radiation from the massive star heats and ionizes the surrounding gas, pushing away the outer layers of the cloud due to radiation pressure. However, the dense inner core of the cloud, which resists ionization, remains largely unaffected. The resulting pressure difference between the outer and inner regions of the cloud leads to the compression of the core, ultimately triggering its collapse and the formation of new stars at the center of the imploding cloud.

While both models incorporate radiation from massive stars, the principal distinction lies in the mechanism of feedback. The C \& C model invokes the expansion of an H\,\textnormal{\textsc{ii}} region that sweeps up and compresses surrounding gas, whereas the RDI model emphasizes the direct action of ionizing radiation on pre-existing dense clumps \citep{porel2025investigating, pandey2013pre}, triggering their collapse. Consequently, the C \& C process typically leads to the formation of clusters with sequential age gradients along swept-up shells \citep{deharveng2005triggered, zavagno2006triggered}, while RDI manifests as bright-rimmed clouds (BRCs) or cometary globules (CGs) at irradiated edges, where star formation occurs in compact cores at the heads of these structures \citep{sugitani1991catalog, miao2009investigation}. Interestingly, \citet{makela2012star} reported evidence of RDI-induced star formation not only at the head but also in the tail region of the CG 1 cloud, suggesting that the influence of ionizing radiation can extend farther into cometary structures than previously assumed. Observationally, these scenarios can be distinguished by morphology and kinematics: C \& C is characterized by large-scale arc or shell-like structures with coherent expansion signatures and young stellar objects (YSOs) distributed along the ionization front, while RDI produces pillar-like or cometary morphologies pointing toward the ionizing source, with embedded YSOs concentrated at their tips. Thus, spatial distributions of YSOs, velocity patterns in molecular tracers, and the geometry of ionized boundaries provide key diagnostics for disentangling the two mechanisms.

Bright Rimmed Clouds are dense molecular clouds that exhibit a prominent, bright edge or rim facing a nearby hot, young massive star or a star-forming region. This bright rim is a result of the interaction between the cloud and the radiation from the nearby star, which ionizes and heats the gas at the cloud’s interface. The ionization and heating cause the gas at the edge of the cloud to become excited, often resulting in observable emission in the form of H$\alpha$, [O III], or other spectral lines. These features are what give BRCs their distinctive bright rims.

SFO 38 belongs to the type-B category of BRCs, distinguished by a sharply curved rim morphology at its head that gradually broadens toward the tail \citep{de2002star}, yielding an overall elongated structure with a length-to-width ratio of 0.68 \citep{sugitani1991catalog}. It is associated with the embedded infrared source IRAS 21391+5802 \citep{sugitani1991catalog}, and lies within the Cepheus OB2 association. This region forms part of a larger ensemble of BRCs linked to active star formation, shaped by the influence of external ionizing radiation from nearby massive stars.
SFO 38, also known as IC 1396N, is located at an approximate distance of 750 pc from the Sun \citep{matthews1979high}. This BRC, alternatively designated based on its location to the north of the H\,\textnormal{\textsc{ii}} region IC~1396, lies within a dynamic environment profoundly shaped by stellar feedback. The H\,\textnormal{\textsc{ii}} region is ionized by the luminous O6.5V-type star HD~206267 \citep{stickland1995spectroscopic, choudhury2010triggered}, a dominant and massive member of the Trumpler~37 open cluster \citep{patel1995large}. The strong UV radiation emitted by HD~206267 sculpts the adjacent interstellar medium, producing a sharply defined ionization front and a characteristic bright rim, indicative of photoionization-driven processes. This radiative influence is understood to induce star formation through the RDI mechanism \citep{sugitani1991catalog}, wherein the advancing ionization front compresses the molecular gas, potentially triggering gravitational collapse. Located at a projected distance of approximately 11~pc from SFO~38, HD~206267 exerts a substantial dynamical and morphological impact on the cloud’s evolution \citep{codella2001star}. Based on astrometric measurements from Gaia Early Data Release 3 (EDR3; \citealt{brown2021gaia}), HD~206267 is located at an estimated distance of approximately 735~pc from the Sun, reinforcing its role as a key ionizing source within the IC~1396 complex. The Figure~\ref{Fig: wise and optical image} displays the WISE 12~$\mu$m intensity map, highlighting the positions of several prominent BRCs—SFO~36, SFO~37, and SFO~39—denoted by red, white, and yellow plus symbols, respectively. A red rectangular box delineates the region associated with SFO~38, which has been extensively mapped using the James Clerk Maxwell Telescope (JCMT) in the $^{12}$CO, $^{13}$CO, and C$^{18}$O J = 3--2 molecular transitions. The white star symbol marks the location of the ionizing star HD~206267.

\cite{serabyn1993pig} conducted a comprehensive analysis of the southern head of the BRC SFO~38, deriving key physical parameters such as density, temperature, mass, and pressure using NH$_{3}$ and CS molecular tracers. Similarly, \cite{sugitani1989star} estimated properties like mass and luminosity for the same region based on observations of the C$^{18}$O J = 1--0 transition, and also reported the detection of molecular outflows associated with the embedded IRAS source in this area. A notable advancement by \cite{beltran2002iras} revealed that the IRAS source IRAS~21391+5802 is resolved into three compact millimeter sources—BIMA~1, BIMA~2, and BIMA~3. Further high-resolution interferometric observations by \citet{neri2007ic1396n} resolved BIMA~2, the most luminous among them, into three distinct components, emphasizing the complex structure of ongoing star formation.

In addition, \cite{codella2001star} reported the discovery of a separate outflow in the northern section of the southern head, reinforcing the region's dynamic nature. \citet{choudhury2010triggered} utilized \textit{Spitzer}-IRAC and MIPS data to identify embedded YSOs and derived a suite of stellar properties including stellar mass, disk mass, effective temperature, and age. The study by \citet{beltran2009stellar} further investigated molecular outflows and the stellar population within the southern head, underlining its active star-forming status.

While extensive work has been carried out on the southern head of SFO~38, the tail regions remain comparatively less explored. \citet{patel1995large} were the first to identify the northwestern (NW) and northeastern (NE) extensions of the cloud—referred to as “wings”—in the $^{12}$CO and $^{13}$CO (J = 1--0) lines, and demonstrated that the NW tail is predominantly blue-shifted, whereas the NE tail exhibits red-shifted kinematics. To date, no definitive evidence of ongoing star formation has been observed in these tail structures.

The star formation observed within the head region of SFO~38, as highlighted in earlier studies, is commonly attributed to the RDI mechanism. However, the extent to which RDI influences the tail regions remains largely unexplored. The primary aim of the present study is therefore to investigate whether a similar RDI-induced star formation scenario operates within the tail regions of SFO~38, by examining both the current star formation activity and the potential for future star formation within these regions. Furthermore, we also consider whether the head region, which is presently active in star formation, is likely to sustain such activity in the future. This approach enables us to obtain a more comprehensive understanding of the role of radiation-driven implosion across the entire cloud. In addition, we perform a comparative analysis of the physical parameters derived for the southern head region with those reported in previous investigations \citep{serabyn1993pig, sugitani1989star}, highlighting discrepancies in parameters such as excitation temperature, mass, and number density estimates, and thereby placing the tail-region findings within the broader context of RDI-driven star formation in SFO~38.

The paper is organized as follows: Section 2 outlines the details of the JCMT archival data used in this study. Section 3 presents the results derived from the data, while Section 4 discusses the analysis based on the gas kinematic data. In Section 5, we delve into the interpretation of the results, and Section 6 provides a summary of the work.


\section{Archive Data} \label{sec:Archive Data}

We used archival data of the $^{12}$CO, $^{13}$CO, and C$^{18}$O $J=3\text{--}2$ transitions obtained with the JCMT, covering an area of approximately $19'\,\times\,16'$ toward the SFO 38 region. These CO isotopologue lines are widely employed as tracers of molecular gas in star-forming regions, providing essential diagnostics of gas density, temperature, and kinematics. The dataset analyzed in this study was acquired on 2008 October 11 as part of the observing program M08BU15. Observations were carried out with the Heterodyne Array Receiver Programme (HARP) in raster scan mode with a scan spacing of 29.10 arcsec. The effective integration times were 12 minutes for $^{12}$CO, and 23 minutes for both $^{13}$CO and C$^{18}$O observations. Data reduction was performed using the ORAC-DR pipeline in the STARLINK software, employing the \texttt{REDUCE\_SCIENCE\_NARROWLINE} recipe \citep{jenness2015orac, buckle2010jcmt}. All datasets have a pixel size of $7.3''$ and velocity channel width of 0.42 km~s$^{-1}$ for $^{12}$CO, and 0.05 km~s$^{-1}$ for both $^{13}$CO and C$^{18}$O, ensuring adequate sampling of the beam and enabling sub-parsec scale structure to be resolved. The mean rms noise levels in the final data cubes are approximately 0.6 K, 1.4 K, and 1.8 K for $^{12}$CO, $^{13}$CO, and C$^{18}$O, respectively. A summary of the observations is provided in Table~\ref{Table: table 1}.

\begin{table*}
\begin{center}
	\caption{Observed CO isotopologue transitions and observational parameters.}
	\label{physical properties of global cloud and clumps}
    \renewcommand{\arraystretch}{1.3} 
	\begin{tabular}{lcccc} 
		\hline
		Transition & Rest frequency & Velocity resolution  & Angular resolution & Mean 1$\sigma$ rms \\ 
		  & (GHz) & (km s$^{-1}$) & ($\arcsec$) & (K)\\ 
		\hline
    $^{12}$CO (3--2)  & 345.79 & 0.42  & 14 & 0.6 \\
    $^{13}$CO (3--2)  & 330.59 & 0.05 & 15 & 1.4 \\
    C$^{18}$O (3--2)  & 329.33 & 0.05 & 15 & 1.8 \\
		\hline 
	\end{tabular}
    \label{Table: table 1}
\end{center}
\end{table*}

\section{Results} \label{sec:Results}

Among the commonly utilized CO isotopologues are $^{12}$CO, $^{13}$CO, and C$^{18}$O, each with unique optical properties suited to different regions within the cloud. The optically thick $^{12}$CO primarily traces the less dense outer regions, while the optically thin C$^{18}$O is adept at probing the dense, inner regions. $^{13}$CO, with its marginally optically thick nature, offers a more comprehensive view by efficiently tracing the entirety of the cloud, providing valuable insights into gas kinematics.

\subsection{Examining Moment Maps and Spectral Features} \label{subsec:Examining Moment Maps and Spectral Features}

\begin{figure*}
\begin{center}
\resizebox{18.0cm}{5.5cm}{\includegraphics{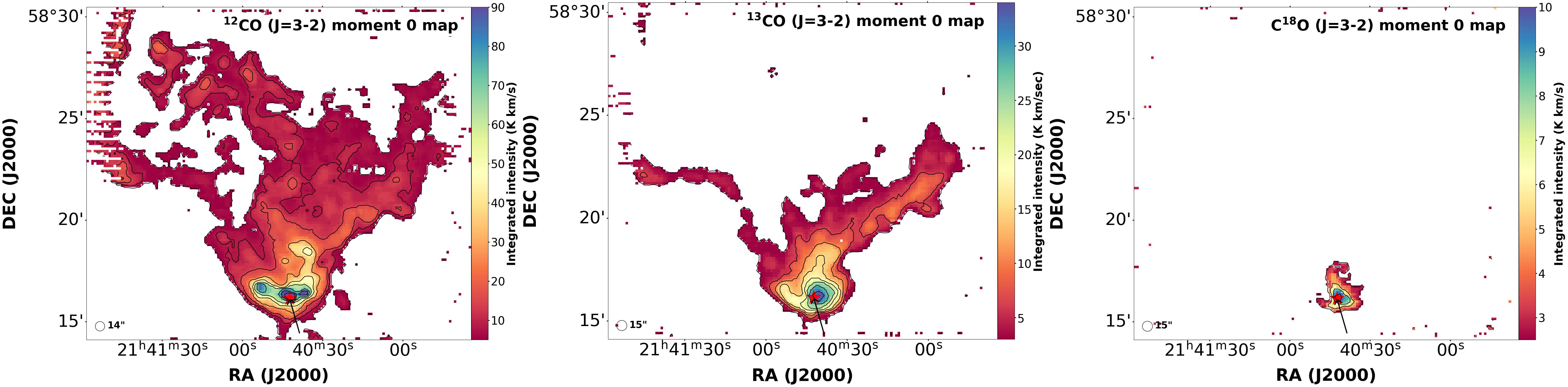}}
\caption{\textbf{Left}: The integrated intensity (moment-0) map of $^{12}$CO (3-2) is presented for the bright-rimmed cloud SFO 38, constructed over the velocity range from -6.38 km s$^{-1}$ to 9.29 km s$^{-1}$. Contours are drawn above the 3$\sigma$ level of the background noise, where $\sigma \approx 1.46$ K km s$^{-1}$ represents the standard deviation of the background noise. The contour levels are set at  3$\sigma$, 6$\sigma$, 10$\sigma$, 20$\sigma$, 30$\sigma$, 40$\sigma$, 50$\sigma$, 60$\sigma$, and 70$\sigma$ values.
\textbf{Middle}: The integrated intensity (moment-0) map of $^{13}$CO (3-2) is shown for cloud SFO 38, created within the velocity range of -2.16 km s$^{-1}$ to 2.44 km s$^{-1}$. Contours are drawn above the 3$\sigma$ threshold of the background noise, where $\sigma \approx 0.71$ K km s$^{-1}$ denotes the standard deviation of the background noise. The contour levels are spaced at intervals of 6$\sigma$.
\textbf{Right}: The integrated intensity (moment-0) map of C$^{18}$O (3-2) is depicted for the cloud SFO 38, created over the velocity range from -1.63 km s$^{-1}$ to 2.21 km s$^{-1}$. The contours are drawn above the 3$\sigma$ level of the background noise, where $\sigma \approx 0.83$ K km s$^{-1}$ represents the standard deviation of the background noise. The contour levels are set at intervals of 3$\sigma$. The red star symbols denote the location of the embedded IRAS source, while the black arrow represents the direction of incident ionizing radiation originating from the massive O-type star HD~206267.}\label{Fig: moment 0 map}
\end{center}
\end{figure*}

Figure~\ref{Fig: moment 0 map} presents the integrated intensity (moment 0) maps of $^{12}$CO (left), $^{13}$CO (middle), and C$^{18}$O (right) emissions. The $^{12}$CO map is constructed over the velocity range of $-6.38$ to $9.29$~km~s$^{-1}$, the $^{13}$CO map spans $-2.16$ to $2.44$~km~s$^{-1}$, and the C$^{18}$O map covers $-1.63$ to $2.21$~km~s$^{-1}$. In all cases, the emission is considered significant only where the intensity exceeds the corresponding mean 1$\sigma$ rms threshold of each molecular tracer. For the $^{12}$CO data, the velocity range was deliberately restricted to exclude the high-velocity blue- and red-shifted components, which are typically associated with molecular outflows \citep{okada2024bright}. The southern part of the cloud exhibits enhanced emission in both $^{12}$CO and $^{13}$CO, while C$^{18}$O, a reliable tracer of dense gas, displays a strong, localized concentration of emission confined to the southern head region. This spatial distribution may be attributed to the RDI mechanism, which leads to the accumulation of dense material in the head of the cloud. Moreover, the $^{12}$CO and $^{13}$CO moment 0 maps demonstrate a systematic decrease in integrated intensity towards the northern portion of the cloud, suggesting a gradient in density or excitation conditions from southern to the northern part of the cloud.

\begin{figure*}
\begin{center}
\resizebox{17.0cm}{12.0cm}{\includegraphics{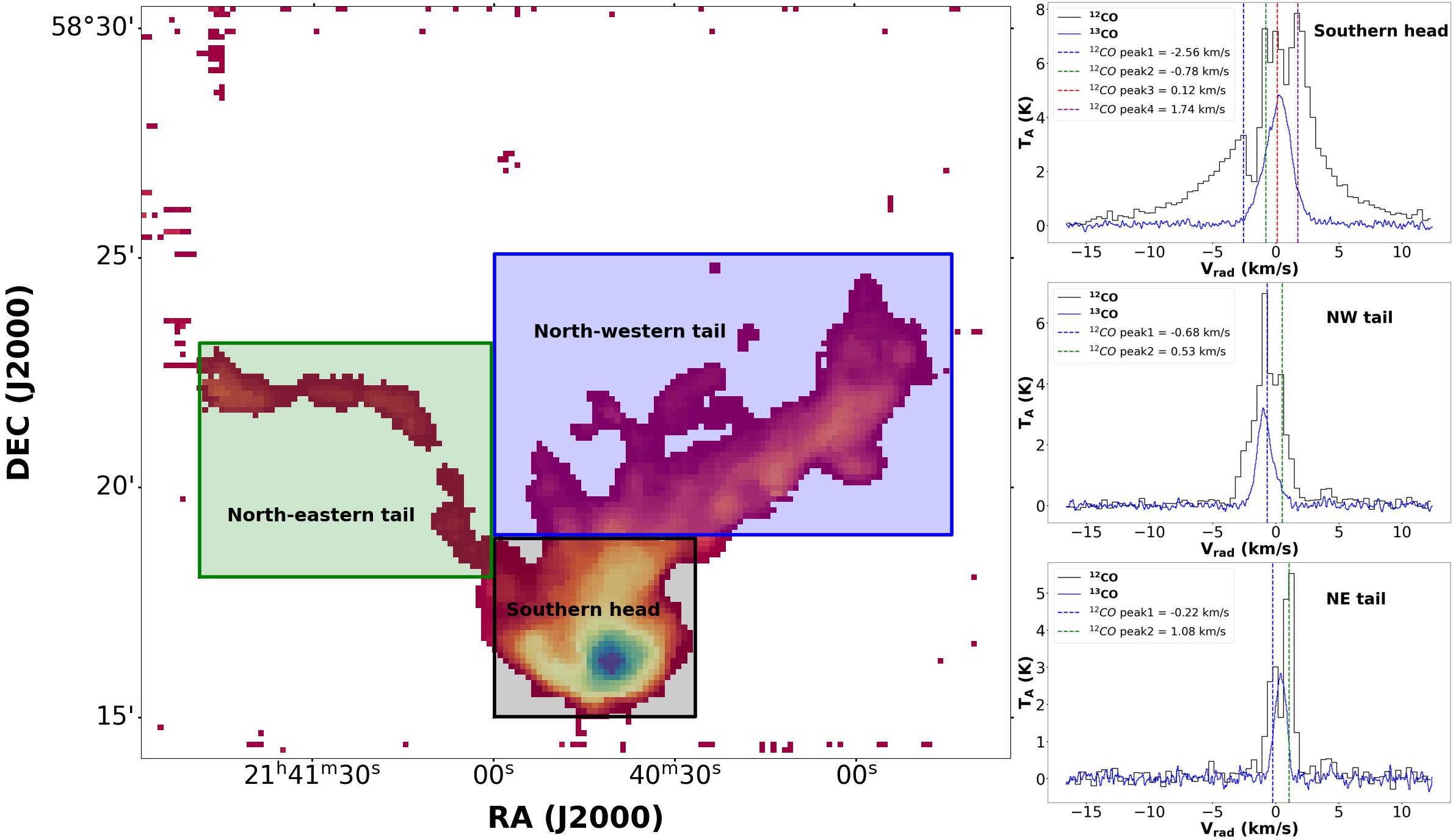}}
\caption{\textbf{Left:} Spatial segmentation of the SFO~38 cloud into three distinct morphological components. The black, blue, and green rectangular overlays on the $^{13}$CO (J=3--2) moment~0 map correspond to the southern head, northwestern tail, and northeastern tail regions, respectively. \textbf{Right:} Spectral profiles extracted from the southern head (top), northwestern tail (middle), and northeastern tail (bottom) are displayed, with $^{12}$CO (J=3--2) and $^{13}$CO (J=3--2) emissions shown in black and blue, respectively. Each $^{12}$CO spectrum is scaled by a factor of 1.5 relative to the $^{13}$CO spectrum, which is baseline-adjusted to 0~K, to enhance visualization and clearly distinguish the spectral profiles of both tracers. \textbf{The vertical dashed lines show different velocity peaks seen in the $^{12}$CO emission.}
}\label{Fig: spectrum of head and tail part}
\end{center}
\end{figure*}

The SFO 38 cloud has been morphologically and kinematically segmented into three distinct regions: the southern head—also identified by \citeauthor{okada2024bright} (\citeyear{okada2024bright}: see their Figure~15)
—along with the north-western (NW) tail and north-eastern (NE) tail also identified by \cite{patel1995large}. These substructures are spatially marked in the left panel of Figure~\ref{Fig: spectrum of head and tail part}. To investigate the internal kinematics, we performed Gaussian fitting on the $^{13}$CO spectral profiles (in blue) corresponding to each region, as shown in the right panel of the same figure. The resulting systemic velocities are 0.23 km s$^{-1}$ for the southern head, –0.87 km s$^{-1}$ for the NW tail, and 0.40 km s$^{-1}$ for the NE tail. The corresponding velocity dispersions, expressed in terms of the full width at half maximum (FWHM, $\Delta V$ = 2.35$\sigma_{obs}$), are measured to be 2.45 km s$^{-1}$, 1.46 km s$^{-1}$, and 1.06 km s$^{-1}$ for the head, NW tail, and NE tail, respectively.

The $^{12}$CO spectral profile (in black) of the southern head reveals a complex velocity structure consisting of four components at –2.56 km s$^{-1}$, –0.78 km s$^{-1}$, 0.12 km s$^{-1}$, and 1.74 km s$^{-1}$, all derived through Gaussian fitting. These components can be formed due to self-absorption by foreground cool components due to the  high optical depth of $^{12}$CO line. Among these, the component at 1.74 km s$^{-1}$ is the most dominant, indicating that red-shifted gas motion prevails in this region. Additionally, the presence of broad wing features in the spectral profile strongly indicates molecular outflow activity, likely driven by ongoing star formation activity in this part of the cloud, is consistent with findings from previous studies. The $^{12}$CO spectra of the NW and NE tails also show two velocity components each, identified through Gaussian decomposition: –0.68 km s$^{-1}$ and 0.53 km s$^{-1}$ in the NW tail (with the former dominating), and –0.22 km s$^{-1}$ and 1.08 km s$^{-1}$ in the NE tail (with the latter being dominant).

These kinematic characteristics suggest that both the southern head and NE tail are predominantly red-shifted relative to their respective systemic velocities, whereas the NW tail is blue-shifted. This pattern is consistent with the findings of \citet{patel1995large}, who discussed similar trends in the SFO 38 tail regions based on velocity channel map analysis. Importantly, our study explicitly determines the systemic velocity of each region, thereby offering a more detailed view of the internal velocity structure of the cloud.

\begin{figure*}
\begin{center}
\resizebox{16.0cm}{6.5cm}{\includegraphics{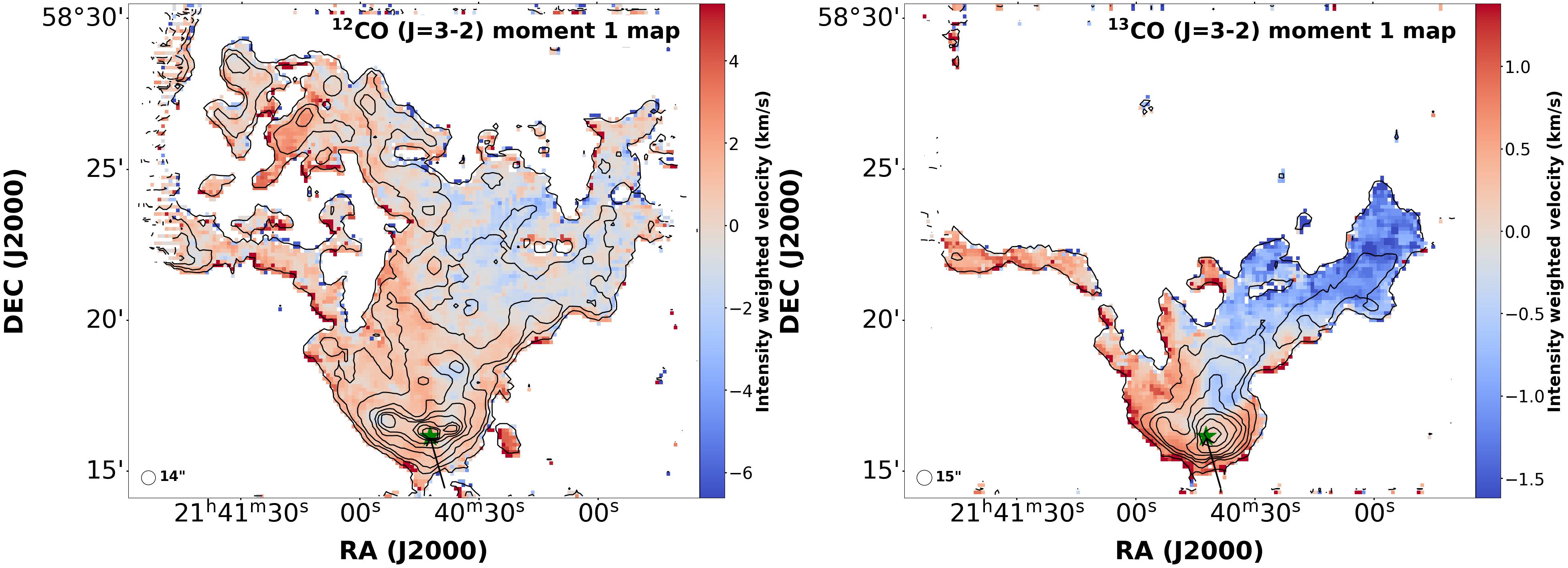}}
\caption{\textbf{Left}: The intensity-weighted velocity (moment-1) map for $^{12}$CO (3–2) emission, constructed over the velocity range from $-6.38$ km s$^{-1}$ to $9.29$ km s$^{-1}$, is displayed. This map is superimposed with the $^{12}$CO integrated intensity contours of the bright-rimmed cloud SFO 38. The contour levels are set to be the same as those in Figure~\ref{Fig: moment 0 map}.  
\textbf{Right}: The intensity-weighted velocity (moment-1) map for $^{13}$CO (3–2) emission, generated within the velocity range of $-2.16$ km s$^{-1}$ to $2.44$ km s$^{-1}$, is presented. Overlaid are the $^{13}$CO integrated intensity contours for the SFO 38 cloud, with contour levels set to be the same as those in Figure~\ref{Fig: moment 0 map}. The green star symbols denote the location of the embedded IRAS source, while the black
arrow represents the direction of incident ionizing radiation originating from the massive O-type star HD 206267.}\label{Fig: moment 1 map}
\end{center}
\end{figure*}

Figure \ref{Fig: moment 1 map} presents the intensity-weighted velocity (moment 1) maps for $^{12}$CO (3–2) (left), and $^{13}$CO (3–2) (right) emission in the SFO 38 cloud. These maps were constructed over the same velocity range used to generate their respective integrated intensity maps. Each moment 1 map is overlaid with the corresponding integrated intensity contours, as shown in Figure \ref{Fig: moment 0 map}.

A well-defined velocity gradient is evident in the $^{13}$CO moment~1 map, particularly from the NW tail to the southern head, reflecting a shift from blue- to red-shifted velocities. This gradient is absent along the NE tail to the southern head, as both regions exhibit predominantly red-shifted velocities.
According to \cite{choudhury2010triggered} (see their figures 1 and 3), analysis of WISE and H$\alpha$ images suggests that the southern head and eastern portion of the SFO 38 cloud are exposed to stronger ionizing radiation from the nearby star HD~206267 than the western side. This anisotropic irradiation may be responsible for imparting greater momentum to the gas in the southern head and NE tail, thereby contributing to their observed red-shifted kinematics relative to the NW tail. However, the observed velocity asymmetry might also be influenced by projection effects associated with the orientation of the individual substructures along the line of sight, and should therefore be interpreted with caution.

\section{Analysis}

In this section, we derive the physical properties of the southern head, the embedded dense clump - defined here as a compact, high column density substructure identified in the C$^{18}$O emission (right panel of Figure~\ref{Fig: moment 0 map}) in this head region, and the NW and NE tail regions of SFO~38 on a pixel-by-pixel basis. The excitation temperature characterizes the thermal excitation conditions within each region, from which we estimate the column density using corresponding optical depth values. These column densities are then used to compute the hydrogen gas mass. The virial parameter analysis assesses the gravitational boundedness of the head and clump regions, offering insight into their potential for future star formation. Additionally, the turbulence parameters, particularly the effective velocity dispersion, provide crucial information on the dynamical state and gravitational stability of the elongated tail structures, helping to evaluate their star-forming potential.

\subsection{Physical Properties of the Southern head}
\label{section: southern head}

\subsubsection{Excitation temperature and optical depth}
\label{section: tex and tau for head}

When the number density of a molecular tracer approaches its critical density, the molecular gas can be considered to be in local thermodynamic equilibrium (LTE). Under LTE conditions, the populations of the molecular energy levels are governed by the Boltzmann distribution, such that the excitation temperature ($T_{\rm ex}$) becomes equivalent to the kinetic temperature ($T_{\rm kin}$) of the gas.  

The excitation temperature of the optically thick $^{12}$CO $J = 3 \rightarrow 2$ transition was derived following the formalism of \citet{buckle2010jcmt}:  
\begin{equation}
    T_{ex}^{12} = \frac{16.59}{\ln \left[ 1 + \frac{16.59}{T_{B_{0}} + 0.036} \right]}.
    \label{eq: 12co tex}
\end{equation}  

In regions affected by self-absorption—identified where the $^{13}$CO peak brightness exceeded that of $^{12}$CO—$T_{\rm ex}$ was instead estimated from the $^{13}$CO transition:  
\begin{equation}
    T_{ex}^{13} = \frac{15.89}{\ln \left[ 1 + \frac{15.89}{T_{B_{0}} + 0.044} \right]}.
    \label{eq: 13co tex}
\end{equation}  
The peak brightness temperature $T_{B_{0}}$ was obtained from the peak antenna temperature using $T_{B_{0}} = T_{A_{0}}/\eta_{mb}$, adopting $\eta_{mb} = 0.61$ for all tracers \citep{buckle2010jcmt}.

\begin{figure*}
\begin{center}
\resizebox{18.0cm}{15.0cm}{\includegraphics{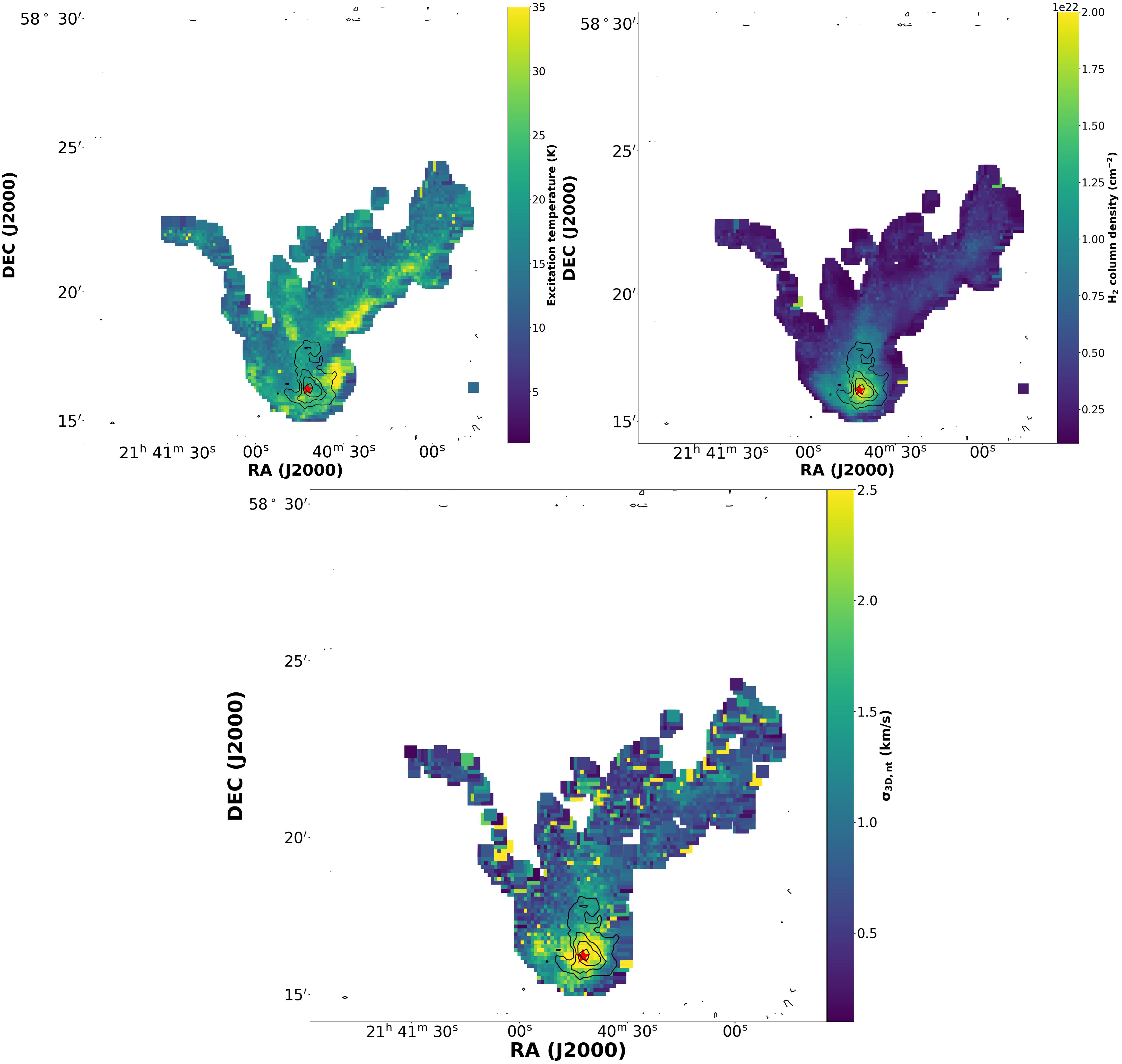}}
\caption{\textbf{Upper left:}Excitation temperature map within $^{13}$CO (3--2) emission region in the SFO 38 bright-rimmed cloud. \textbf{Upper right:} Hydrogen column density map based on $^{13}$CO emission. \textbf{Lower:} 3-dimensional non-thermal velocity dispersion map based on $^{13}$CO emission. Each map is overlaid with the C$^{18}$O integrated intensity contours. The red star symbols show the position of the IRAS source.}\label{Fig: tex cd sigma3nt map}
\end{center}
\end{figure*}

The upper left panel of Figure~\ref{Fig: tex cd sigma3nt map} showcases the spatial distribution of excitation temperature derived by utilizing equations (\ref{eq: 12co tex}) and (\ref{eq: 13co tex}). The mapped region is confined within the outermost contour, defined by an integrated intensity threshold of 3$\sigma$ in $^{13}$CO emission.

For the southern head region, the excitation temperature lies within 6.2--41.5 ~K, with a mean value of approximately 20.6~K, and within the clump, the corresponding values are 15.7--33.9 K with a mean value around 22.5 K. In comparison, \citet{serabyn1993pig} derived excitation temperatures from NH$_3$ (1,1) and (2,2) inversion transitions, reporting a maximum value of 26~K near the southern ionization front and values between 20--23~K in the vicinity of the associated IRAS source. Our mean excitation temperature, derived under the assumption of local thermodynamic equilibrium, are in excellent agreement with their results. The higher maximum temperature observed in our study is expected, as our excitation temperatures are estimated from the optically thick $^{12}$CO and the marginally optically thick $^{13}$CO ($J=3\text{--}2$) transitions, which predominantly trace the warmer, more diffuse outer layers of the molecular cloud exposed to stronger ionizing radiation. In contrast, NH$_3$ traces the colder, denser inner regions, being a high-density tracer less influenced by the external radiation field. This distinction highlights the stratified thermal structure of the cloud, shaped by both radiative heating and density-dependent excitation conditions. Table~\ref{Table: table 3} summarizes the range and mean values of the excitation temperature for the southern head part.

The optical depth equations for $^{13}$CO (3-2) and C$^{18}$O (3-2) emission are determined using the following expressions::

\begin{equation}
    \tau_{0}^{13} = -\ln\left(1 - \frac{T^{13}_{B_{0}}}{15.89}\left[\frac{1}{\exp\left(\frac{15.89}{T_{ex}}\right) - 1} - \left(2.79 \times 10^{-3}\right)\right]^{-1}\right).
    \label{eq: 13co tau equation}
\end{equation}

\begin{equation}
    \tau_{0}^{18} = -\ln\left(1 - \frac{T^{18}_{B_{0}}}{15.79}\left[\frac{1}{\exp\left(\frac{15.79}{T_{ex}}\right) - 1} - \left(2.89 \times 10^{-3}\right)\right]^{-1}\right).
    \label{eq: c18o tau equation}
\end{equation}

Using Equation~\ref{eq: 13co tau equation}, we derive the optical depth of the $^{13}$CO (3--2) emission to range from 0.02 to 3.57, with a mean value of approximately 1.06. Within the clump region, the optical depth of the C$^{18}$O (3--2) line---calculated using Equation~\ref{eq: c18o tau equation}---varies between 0.03 and 0.99, with a mean value around 0.51. These results suggest that the $^{13}$CO (3--2) emission in the southern head of SFO 38 is generally in the regime of moderate optical thickness, while the C$^{18}$O (3--2) emission remains largely optically thin. Table~\ref{Table: table 2} gives the mean value of the optical depth in $^{13}$CO emission for the southern head part. 

\subsubsection{Mass and stability analysis}
\label{section: mass and stability}

Given that $^{12}$CO is a highly optically thick tracer, it primarily traces the warmer and more diffuse outer envelopes of molecular clouds. In contrast, $^{13}$CO, with its comparatively moderate optical depth, serves as a more reliable tracer of the bulk molecular gas distribution. Consequently, to better characterize the physical conditions within the southern head of the SFO 38 bright-rimmed cloud, we compute the $^{13}$CO column density across the broader southern head region, and the C$^{18}$O column density specifically within the clump identified in the C$^{18}$O (3--2) emission map. These column densities were estimated using the formulation presented by \citet{hayashi1991rp}, which is appropriate under the assumption of local thermodynamic equilibrium.

\begin{equation}
    N\left(^{13}\mathrm{CO}\right) = 8.28 \times 10^{13} \exp\left(\frac{15.87}{T_{\mathrm{ex}}}\right) 
    \times \frac{T_{\mathrm{ex}} + 0.88}{1 - \exp\left(-\frac{15.87}{T_{\mathrm{ex}}}\right)} \int \tau\,dv,
    \label{eq: 13co column density}
\end{equation}

\begin{equation}
    N\left(C^{18}O\right) = 8.26 \times 10^{13} \exp\left(\frac{15.81}{T_{ex}}\right) 
    \times \frac{T_{ex} + 0.88}{1 - \exp\left(-\frac{15.81}{T_{ex}}\right)} \int \tau\,dv,
    \label{eq: c18o column density}
\end{equation}

where the integrated optical depth, $\int \tau\,dv$, is connected to the observed brightness temperature via:

\begin{equation}
    \int \tau\,dv = \frac{1}{\left[J\left(T_{\mathrm{ex}}\right) - J\left(T_{\mathrm{bg}}\right)\right]} \int T_{\mathrm{B}}\,dv \quad \text{for } \tau < 1,
    \label{eq: tau < 1 condition}
\end{equation}

and in optically thick regimes ($\tau \geq 1$), the correction becomes:

\begin{equation}
    \int \tau\,dv = \frac{1}{\left[J\left(T_{\mathrm{ex}}\right) - J\left(T_{\mathrm{bg}}\right)\right]} \cdot \frac{\tau}{1 - e^{-\tau}} \int T_{\mathrm{B}}\,dv.
    \label{eq: tau > 1 condition}
\end{equation}

Subsequently, the H$_2$ column density can be estimated from the $^{13}$CO and C$^{18}$O  column density using the conversion relation \textbf{\citep{frerking1982relationship, rawat2024giant}}:

\begin{equation}
    \left(N\left(\mathrm{H}_{2}\right)\right)_{^{13}\mathrm{CO}} = 7 \times 10^{5} \, N\left(^{13}\mathrm{CO}\right),
    \label{eq: h2 column density relation}
\end{equation}
and
\begin{equation}
    \left(N\left(H_2\right)\right)_{C^{18}O} = 7 \times 10^{6} \, N\left(C^{18}O\right).
    \label{eq: h2 cd based on c18o}
\end{equation}

The upper right panel of Figure~\ref{Fig: tex cd sigma3nt map} illustrates the spatial distribution of the H$_2$ column density within the region enclosed by the outermost (3$\sigma$) contour of the $^{13}$CO (3--2) emission based on this molecular tracer. Overlaid upon this map are the C$^{18}$O integrated-intensity contours (black), which serve to trace the denser substructures. The highest column density aligns with the position of the embedded IRAS source, supporting the scenario of ongoing or imminent star formation within a dense environment.

The total mass of the molecular cloud was estimated using the following relation:
\begin{equation}  
    M = \mu_{H_{2}}\, m_{H}\, A_{\mathrm{pixel}} \sum N\left(\mathrm{H}_{2}\right),
    \label{eq: mass relation}
\end{equation}
where $\mu_{H_{2}}$ is the mean molecular weight per hydrogen molecule, adopted as 2.8 following \citet{kauffmann2008mambo}, $m_{H}$ is the mass of a hydrogen atom, $A_{\mathrm{pixel}}$ denotes the physical area of a single pixel in units of cm$^{2}$. 

For the mass estimation of the southern head of the SFO 38 bright-rimmed cloud, we employed a two-component approach: the H$_2$ column density derived from C$^{18}$O emission was used within the dense clump region, while the $^{13}$CO-based column density was adopted for the surrounding lower-density areas. Applying this methodology, the total mass of the southern head was found to be approximately $204$~M$_\odot$.

Assuming circular geometries for both the C$^{18}$O clump and the broader southern head within the SFO 38 cloud, we derived the effective radii using $r_{\rm eff} = \sqrt{A / \pi}$, where A denotes the respective projected areas. Based on this approach, the calculated effective radii for the dense clump and the entire southern head region are approximately 0.22~pc and 0.47~pc, respectively.

\citet{sugitani1989star} estimated the mass of the globule to be approximately 150~M$_\odot$, assuming a radius between 0.1 and 0.2~pc, based on $^{13}$CO (1--0) observations carried out with the NRO 45-m telescope, which provided an angular resolution of 17$\arcsec$ at a wavelength of 2.6~mm. In comparison, our analysis yields a clump mass of $\sim$90~M$_\odot$ within a radius of 0.22~pc based on the C$^{18}$O (3--2) transition, and $\sim$70~M$_\odot$ from the $^{13}$CO (3--2) transition. The discrepancy between our mass estimate from C$^{18}$O (3--2) and that reported by \citet{sugitani1989star} from $^{13}$CO (1--0) may arise from the higher abundance and lower self-absorption characteristics of the $^{13}$CO (1--0) transition, which, coupled with its ability to trace more extended diffuse gas, can lead to elevated mass estimates.

Additionally, \citet{serabyn1993pig} reported a significantly higher mass of $\sim$480~M$_\odot$ for the globule within a radius of 0.29~pc using CS line observations from the IRAM 30-m telescope. This elevated mass estimate can be attributed to the fact that CS is a tracer of much higher density gas, typically probing regions with densities in the range of $10^4$ to $10^7$~cm$^{-3}$ \citep{serabyn1993pig}, compared to the C$^{18}$O (3--2) transition, which is sensitive to densities of approximately $10^4$ to $10^5$~cm$^{-3}$. Therefore, the use of CS enables the detection of deeply embedded, compact high-density cores that may be underrepresented in C$^{18}$O-based estimates, leading to a larger inferred mass in their study.

The lower mass of the clump derived from $^{13}$CO (3--2) in our study, relative to that from C$^{18}$O (3--2), can be attributed to the effects of self-absorption within the clump, which leads to an underestimation of the column density. This supports the interpretation that C$^{18}$O (3--2), being optically thin and less susceptible to self-absorption, serves as a more reliable tracer for clump mass estimation in dense environments.

Assuming a spherical geometry for the clump as well as for the southern head, we estimate the volume number density of molecular hydrogen (\( n_{\mathrm{H_2}} \)) using the following expression:

\begin{equation}
    n_{\mathrm{H_2}} = \frac{3M}{4\mu_{\mathrm{H_2}}\, m_{\mathrm{H}}\, \pi\, r_{eff}^{3}},
\end{equation}

Using this formalism, we estimate the molecular hydrogen number density ($n_{\mathrm{H}_2}$) within the C$^{18}$O clump to be approximately $3.0 \times 10^{4}$~cm$^{-3}$, while the average number density across the entire southern head region is found to be around $6.8 \times 10^{3}$~cm$^{-3}$. These values are consistent with earlier results reported by \citet{sugitani1989star}, who derived a number density of $\sim5 \times 10^{4}$~cm$^{-3}$, in good agreement with our clump-scale estimate.

On the other hand, \citet{serabyn1993pig} reported a significantly higher number density of $\sim8 \times 10^{5}$~cm$^{-3}$ based on observations of the optically thin C$^{34}$S (5--4) and (2--1) transitions. This elevated density estimate is expected, as C$^{34}$S is a tracer of much denser gas compared to C$^{18}$O. While both are optically thin, C$^{34}$S transitions probe regions with substantially higher critical densities, typically associated with compact, deeply embedded core structures that are not fully traced by C$^{18}$O emission.

To assess the stability of the C$^{18}$O clump and the southern head within SFO 38 cloud under gravitational influences, we performed a virial analysis. The virial mass is derived from the following relation \citep{maclaren1988corrections, rawat2023probing}: 
\begin{equation}  
    M_{vir} = 126 \left(\frac{5 - 2\beta}{3 - \beta}\right) \left(\frac{r_{eff}}{pc}\right) \left(\frac{\Delta V}{\mathrm{km\,s^{-1}}}\right)^{2},
    \label{eq: virial mass}
\end{equation}  
where we have adopted $\beta = 2$ assuming the density profile as $\rho \propto r^{-\beta}$ \citep{maclaren1988corrections, rawat2023probing}. The $\Delta V$ entering the virial mass calculation was estimated differently for the clump and for the broader southern head. Within the clump boundary, $\Delta V$ was derived exclusively from the pixel-averaged C$^{18}$O (3--2) line widths, since this optically thin tracer reliably probes the dense gas component directly associated with the clump. For the larger southern head region, $\Delta V$ was determined by combining C$^{18}$O (3--2) line widths inside the clump with $^{13}$CO (3--2) line widths outside the clump, thereby capturing both the dense and more extended gas components. We emphasize that $\Delta V$ was computed as the average of individual pixel line widths, rather than being extracted from the line profile of an average spectrum of each region. This approach minimizes the risk of artificially broadening the line width due to averaging over regions with different systemic velocities or complex kinematics. By considering the local FWHM at each pixel and then averaging, we obtain a more accurate representation of the internal motions within the clump and head, thereby ensuring that the virial analysis reflects the true dynamical state of the system. Based on this methodology, the virial masses of the C$^{18}$O clump and the entire southern head region of the SFO 38 cloud are estimated to be 86~M$_\odot$ and 162~M$_\odot$, respectively. The virial parameter $\alpha$, which quantifies the cloud's gravitational stability, is calculated using $\alpha = M_{\rm vir}/M$. For the clump and the southern head of the cloud, the calculated virial parameters are 0.95 and 0.79, respectively. Since both values are below the critical threshold of 2, this implies that the clump and the larger southern head structure are gravitationally bound. Such boundedness is consistent with the observed signatures of active star formation in the region, including the presence of a young stellar population \citep{choudhury2010triggered, beltran2009stellar} and multiple molecular outflows linked to protostellar activity \citep{beltran2002iras, neri2007ic1396n, codella2001star, fuente2009dissecting, okada2024bright}, particularly prominent in the southern portion of the cloud.

Table~\ref{Table: table 3} presents the minimum, maximum, and mean values of $N(\mathrm{H}_2)$, effective radius, mass, $n(\mathrm{H}_2)$ and gravitational stability of the southern head part, providing a comprehensive overview of the physical conditions across the region.

\subsubsection{Turbulence parameters}

To investigate the turbulent characteristics within the southern head as well as the clump of the SFO 38 cloud, we derived several key parameters, including the thermal velocity dispersion, non-thermal velocity dispersion, and Mach number. The thermal velocity dispersion (\(\sigma_{th}\)) is computed using the expression:

\begin{equation}
    \sigma_{th} = \sqrt{\frac{k_B T}{\mu_i m_H}},
    \label{eq: thermal velocity dispersion}
\end{equation}
where the temperature $T$ is taken to be the excitation temperature obtained from the region as traced by $^{13}$CO, \(\mu_i\) denotes the molecular mass of the tracers, with values of 29 for $^{13}$CO, and 30 for C$^{18}$O \citep{rawat2024giant}.

The one-dimensional non-thermal velocity dispersion (\(\sigma_{nt, 1D}\)) is given by:

\begin{equation}
    \sigma_{nt, 1D} = \sqrt{\sigma_{obs}^2 - \sigma_{th}^2}.
    \label{eq: non thermal 1d velocity dispersion}
\end{equation}

The three-dimensional non-thermal velocity dispersion (\(\sigma_{nt, 3D}\)) is subsequently derived as $\sigma_{nt, 3D} = \sqrt{3} \sigma_{nt, 1D}$.

The Mach number (\(\mathcal{M}\)), an indicator of turbulence, is defined as:

\begin{equation}
    \mathcal{M} = \frac{\sigma_{nt, 3D}}{c_s}.
    \label{eq: Mach number}
\end{equation}
The thermal sound speed ($c_{\rm s}$) is calculated with $c_{\rm s}=\sqrt{k_{\rm B} T / \mu m_{\rm H}}$, where $\mu$ is the mean molecular weight per free particle of the gas, taken as 2.37 \citep{kauffmann2008mambo}.

The lower panel of Figure~\ref{Fig: tex cd sigma3nt map} depicts the spatial distribution of $\sigma_{nt, 3D}$, across the entire SFO 38 cloud as traced by $^{13}$CO emission. An enhanced non-thermal velocity dispersion is observed in the southern head region, particularly in the vicinity of the IRAS source, which is likely attributed to energetic feedback processes such as molecular outflows driven by the source. The derived mean Mach numbers of the clump and the entire southern head are nearly identical, 4.65 and 4.66, respectively, using the same approach as calculation of virial mass, indicating a supersonic regime throughout the southern head. These elevated Mach numbers and velocity dispersions collectively signify that the southern head is highly turbulent, driven by the synergistic effects of RDI and ongoing stellar feedback from young stellar objects. Table~\ref{Table: table 2} summarizes these turbulent parameters of the southern head of SFO 38 cloud.

\subsection{Physical Properties of NW and NE tails}
\label{section: NW and NE tails}

In this section, we examine the physical properties of the NW and NE tails in a similar manner to the southern head.

\subsubsection{Excitation temperature and optical depth}
\label{section: tex and tau for tails}

For the NW tail, the excitation temperature spans 6.3--39.5~K, with a mean value of 17.5~K. In the NE tail, it ranges from 6.2--41.5~K, with a mean of 14.9~K. Table~\ref{Table: table 3} summarises the range and mean values of the excitation temperature for these tail regions. A striking feature emerges in the western segment of the NW tail, as shown in the upper left panel of Figure~\ref{Fig: tex cd sigma3nt map}, where significantly elevated excitation temperatures are observed compared to the rest of the NW tail and the entire NE tail.

Although this part of the NW tail experiences relatively less intense ionizing radiation from the star HD~206267, it shows elevated excitation temperatures in the $^{13}$CO emission. This suggests that local shielding may play a key role in trapping thermal energy and sustaining higher excitation temperatures.

Conversely, the NE tail, despite being directly exposed to the ionization front of HD~206267, displays comparatively lower excitation temperatures. This contrast can be attributed to the destructive effects of strong ionizing radiation, which can drive photoevaporation of molecular gas and suppress molecular excitation, particularly in the $^{13}$CO-emitting layers, leading to a cooler and more diffuse gas phase.

Based on the derived excitation temperatures for the southern head, NW tail, and NE tail, the mean value indicate that the NE tail exhibits the lowest excitation temperature, despite being exposed to significant UV radiation. This suggests that the elevated excitation temperature observed in the southern head is not solely attributed to RDI, but is also influenced by active protostellar processes evident in that region. 

Using Equation~\ref{eq: 13co tau equation}, the optical depth ($\tau$) of the $^{13}$CO (3--2) emission was derived for both the NW and NE tails. In the NW tail, the optical depth spans 0.06--3.95, with a mean value of approximately 1.09. Similarly, in the NE tail, $\tau$ ranges from 0.14--3.33, yielding a mean value of about 1.03. These values indicate that the $^{13}$CO (3--2) transition is predominantly optically thin to moderately thick across both regions, with comparable average opacities. Table~\ref{Table: table 2} shows the mean values of the $^{13}$CO optical depth for the tail regions.

\subsubsection{Mass and stability analysis}
\label{section: mass of tail}

The upper right panel of Figure~\ref{Fig: tex cd sigma3nt map} presents the spatial distribution of the derived H$_2$ column density across the NW and NE tails. It is evident from the map that both tails exhibit comparatively lower column densities than the head region, a characteristic trend expected in regions shaped by the RDI mechanism.

Using the equation~\ref{eq: mass relation}, the mass of the NW tail and NE tail within the SFO 38 bright-rimmed cloud has been estimated to be 135~M$_\odot$ and 19~M$_\odot$, respectively, based on $^{13}$CO (3--2) emission.

Following the method proposed by \citet{stobie1980application} and adopting the formulation of \citet{patel1995large}, the parameters of the ellipses fitted to the NW and NE tails were derived from the second-order moments of the intensity distribution. In this formalism, the semi-major axis $a$, semi-minor axis $b$, and position angle $\phi$ of the major axis of the best-fit ellipse measured from the east are obtained directly from the intensity-weighted variances and covariance of the pixel coordinates. The variances along the $x$ and $y$ directions, together with their covariance, are computed with respect to the intensity-weighted centroid of the structure, ensuring that the fit accounts for the spatial distribution of emission. Applying this method, we obtained projected major and minor axes of 1.92~pc and 1.10~pc for the NW tail, and 1.77~pc and 1.10~pc for the NE tail. The corresponding position angles of the NW tail and NE tail are approximately $151^\circ$ and $24^\circ$. Notably, despite the minor axes being nearly identical, the visual morphology in Figure~\ref{Fig: moment 0 map} clearly shows structural asymmetry between the two tails, particularly in their widths. This mismatch suggests that the moment-based ellipse-fitting approach may underestimate differences in transverse extent, making it less reliable for capturing such asymmetries.

To obtain a more physically meaningful estimate of the tail widths, we therefore adopt an elliptical geometric approach based on the projected area. Specifically, we define the width \(W\) as: $W = 2\left(\frac{A}{\pi a}\right)$ where \(A\) is the projected area and \(L = 2a\) is the full length of the tail derived from the major axis. Using this method, we find the widths of the NW and NE tails to be approximately 0.70~pc and 0.19~pc, respectively—values that are more consistent with the morphology observed in the moment 0 map and thus considered more reliable representations of the tails' physical widths.

Assuming a cylindrical geometry for the NW and NE tails, we estimate the volume number density of molecular hydrogen ($n_{\mathrm{H}_2}$) using the following expression:
\begin{equation}
    n_{\mathrm{H}_2} = \frac{M}{\mu_{\mathrm{H}_2} \, m_{\mathrm{H}} \, \pi r^2 L},
    \label{eq: volume_density}
\end{equation}
where $L$ is the length of the cylinder, and $r = W/2$ denotes the radius, with $W$ being the width of the tail.

Using this formulation, we derive volume number densities of $n_{\mathrm{H}_2} \approx 2.7 \times 10^{3}$~cm$^{-3}$ for the NW tail and $\approx 5.3 \times 10^{3}$~cm$^{-3}$ for the NE tail, indicating that both structures are composed of moderately dense molecular gas. These values of $n_{H_{2}}$ are below the typical threshold ($\geq 10^{4}-10^{5}$ cm$^{-3}$) required for spontaneous gravitational collapse and the formation of stars \citep{bergin2007cold, mckee2007theory}.


To assess the dynamical stability of these tail regions, we examine two key parameters: the \textit{line mass} (\( M_{\text{line}} \)) and the \textit{critical line mass} $M_{\rm line}^{\rm crit}$, following the formalism presented in \citet{fiege2000helical}. These quantities are defined as:

\begin{equation}
    M_{\text{line}} = \frac{M}{L},
\end{equation}

and

\begin{equation}
    M_{\rm line}^{\rm crit} = \frac{2\sigma_{eff}^{2}}{G} \sim 464\, \sigma_{\text{eff}}^2\ M_{\odot}\,\text{pc}^{-1},
\end{equation}

where G is the gravitational constant, and \( \sigma_{\text{eff}} \) is the effective velocity dispersion given by:

\begin{equation}
    \sigma_{\text{eff}} = \sqrt{\sigma_{\text{nt, 1D}}^2 + c_s^2},
\end{equation}

Using the above expressions, we calculate the line masses for the NW and NE tails to be approximately 70 M$_\odot$pc$^{-1}$ and 11 M$_\odot$pc$^{-1}$, respectively. The corresponding critical line masses are found to be 151 M$_\odot$pc$^{-1}$ and 142 M$_\odot$pc$^{-1}$. In both cases, the observed line masses fall significantly below the critical thresholds, indicating that these tail-like structures are gravitationally unbound. This conclusion is further supported by the spectral line profiles, which exhibit signatures of outward motions suggestive of gas dispersal as discussed in section~\ref{section: Spectral Profile-Based Gas Kinematics Study of the Northwestern and Northeastern Tails of SFO 38}. The lack of gravitational binding, coupled with kinematic evidence of expansion, implies that these tails are unlikely to undergo gravitational collapse, thereby inhibiting future star formation within them.

Table~\ref{Table: table 3} presents the minimum, maximum, and mean values of $N(\mathrm{H}_2)$, major and minor axes, mass, volume-averaged density $n(\mathrm{H}_2)$, and the gravitational stability of the NW and NE tail regions.

\begin{table*}
\begin{center}
    \caption{Physical properties and gravitational stability nature of the southern head, NW tail, and NE tail of the SFO 38 cloud.}
    \label{Physical properties for SFO 38}
    \renewcommand{\arraystretch}{1.3} 
    \begin{tabular}{l ccc ccc c c c c} 
        \hline
        Region & \multicolumn{3}{c}{$T_{\rm ex}$ (K)} & \multicolumn{3}{c}{$N(H_2)$ ($\times10^{22}$ cm$^{-2}$)} & size$^{\dagger}$ & Mass & $\langle n(H_2)\rangle$ & Stability \\
        \cline{2-4} \cline{5-7}
               & Min & Max & Mean & Min & Max & Mean & (pc) & ($M_\odot$) & ($\times10^3$ cm$^{-3}$) & \\
        \hline
        Southern head & 6.2 & 41.5 & $20.6\pm0.2$ & 0.2 & 3.4 & $1.2\pm0.4$ & 0.47 & $204\pm17$ & $6.8\pm0.6$ & Unstable\\
        NW tail & 6.3 & 39.5 & $17.5\pm0.1$ & 0.5 & 2.5 & $0.6\pm0.1$ & 1.92 $\times$ 0.70 & $135\pm21$ & $2.7\pm0.4$ & Stable \\
        NE tail & 6.2 & 41.5 & $14.9\pm0.3$ & 0.2 & 2.8 & $0.5\pm0.1$ & 1.77 $\times$ 0.19 &  $19\pm3$ & $5.3\pm1.3$ & Stable\\
        \hline 
    \end{tabular}
    \tablecomments{$^{\dagger}$ The size of the southern head is presented as the radius, whereas for the NW and NE tails it is expressed as length $\times$ width. For further details, see the text.}
    \label{Table: table 3}
\end{center}
\end{table*}

\begin{table*}
\begin{center}
	\caption{Kinematic and turbulent properties of the southern head, NW tail, and NE tail in the SFO 38 bright-rimmed cloud.}
	\label{physical properties of global cloud and clumps}
    \renewcommand{\arraystretch}{1.3} 
	\begin{tabular}{lccccccccc} 
		\hline
		Region & $\langle \tau_{13} \rangle$ & $\Delta V$  & $\sigma_{3D}$ & $V_{LSR}$ & $\langle \sigma_{th} \rangle$ & $\langle \sigma_{nt, 3D} \rangle$ & $\langle c_{s} \rangle$ & $\langle \mathcal{M} \rangle$ \\ 
		  &  & (km s$^{-1}$) & (km s$^{-1}$) & (km s$^{-1}$) & (km s$^{-1}$) & (km s$^{-1}$) & (km s$^{-1}$) & \\ 
		\hline
    Southern head & $1.06\pm0.03$ & $2.44\pm0.02$ & $1.80\pm0.01$ & $0.23\pm 0.01$ & 0.07 & $1.19\pm0.02$ & 0.26 & $4.66\pm0.10$\\
    NW tail & $1.09\pm0.02$ &  $1.46\pm0.02$ & $1.07\pm0.01$ & $-0.87\pm0.01$ & 0.07 & $0.89\pm0.02$ & 0.24 & $3.75\pm0.08$\\
    NE tail & $1.03\pm0.05$ &  $1.06\pm0.02$ & $0.78\pm0.01$ & $0.40\pm0.01$ & 0.06 & $0.87\pm0.05$ & 0.23 & $3.96\pm0.25$\\
		\hline 
	\end{tabular}
    \label{Table: table 2}
\end{center}
\end{table*}

\subsubsection{Turbulent properties}
\label{section: turbulent properties}

The lower panel of Figure~\ref{Fig: tex cd sigma3nt map} shows that in both tails the non-thermal velocity dispersion is lower than the southern head because of the absence of star formation activity within the tail regions. Using the relevant analytical expressions (equations~\ref{eq: thermal velocity dispersion} to~\ref{eq: Mach number}), the mean Mach number, indicative of the level of turbulence relative to the thermal sound speed, is found to be $\sim$3.7 in the NW tail of SFO 38. Similarly, NE tail, the mean Mach number is estimated to be $\sim$4.0, suggesting that supersonic turbulence prevails throughout both tails. Table~\ref{Table: table 2} summarizes these turbulent parameters of the NW and NE tails of SFO 38 cloud.

\subsection{Spectral Profile-Based Gas Kinematics Study of the Northwestern and Northeastern Tails of SFO 38}
\label{section: Spectral Profile-Based Gas Kinematics Study of the Northwestern and Northeastern Tails of SFO 38}

\begin{figure*}
\begin{center}
\resizebox{18.0cm}{17.5cm}{\includegraphics{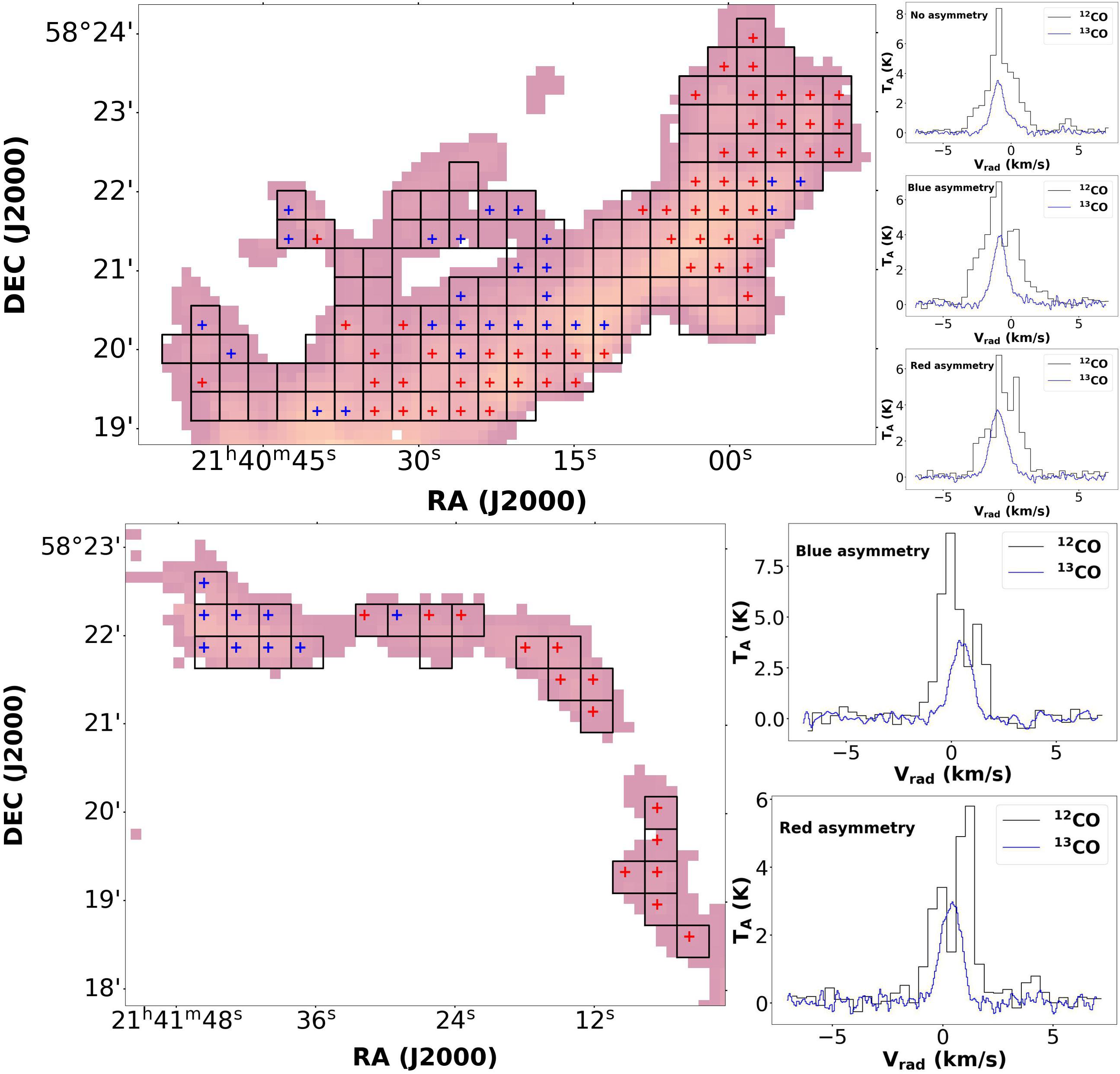}}
\caption{\textbf{Left:} The upper and lower left panels depict box regions arranged in a 3$\times$3 grid, each approximately $22\arcsec \times 22\arcsec$ in size, overlaid on the $^{13}$CO emission maps of the NW and NE tails, respectively. Blue and red plus symbols indicate positions exhibiting significant velocity asymmetries, defined by $\delta V < -0.25$ and $\delta V > 0.25$, respectively, for both tail regions.
\textbf{Right:} The upper right panel presents the averaged spectral profiles of $^{12}$CO (black) and $^{13}$CO (blue) for three categories of regions within the NW tail: symmetric regions without asymmetry (top), blue-asymmetric regions ($\delta V < -0.25$; middle), and red-asymmetric regions ($\delta V > 0.25$; bottom), corresponding to the box selections in the left panel. The lower right panel shows analogous spectral profiles for the NE tail, with blue- and red-asymmetric regions (top and bottom, respectively) identified by blue and red plus symbols. Each $^{12}$CO spectrum is scaled by a factor of 1.5 relative to the $^{13}$CO spectrum, which is baseline-adjusted to
0 K, to enhance visualization and clearly distinguish the spectral profiles of both tracers.
}\label{Fig: infall and expansion}
\end{center}
\end{figure*}

To investigate kinematic signatures indicative of infall motion or expansion within molecular cloud regions, \citet{mardones1997search} introduced a dimensionless asymmetry parameter defined as:
\begin{equation}
    \delta V = \frac{V_{\text{thick}} - V_{\text{thin}}}{\Delta V_{\text{thin}}},
    \label{eq:asymmetry_equation}
\end{equation}
where $V_{\text{thick}}$ and $V_{\text{thin}}$ denote the peak velocities from Gaussian fits to the optically thick ($^{12}$CO) and optically thin ($^{13}$CO) spectral lines, respectively. The quantity $\Delta V_{\text{thin}}$ represents the FWHM of the $^{13}$CO line. A line profile is classified as \textit{blue-asymmetric} (indicative of infall) when $\delta V < -0.25$, \textit{red-asymmetric} (suggestive of expansion) when $\delta V > 0.25$, and \textit{symmetric} when $-0.25 < \delta V < 0.25$.

Figure~\ref{Fig: infall and expansion} displays box-shaped regions, each approximately 0.08\,pc\,$\times$\,0.08\,pc in size, overlaid on the NW tail (upper left) and NE tail (lower left) of the SFO~38 cloud in the $^{13}$CO moment 0 map. Based on their spectral profiles, the regions are categorized as follows: unmarked boxes represent symmetric profiles, blue plus signs indicate blue-asymmetric profiles, and red plus signs denote red-asymmetric features.

In the NW tail, two regions exhibit double-peaked profiles in both $^{12}$CO and $^{13}$CO, but the alignment of their peaks suggests the presence of overlapping velocity components rather than genuine infall. Another four regions show single-peaked profiles in both tracers. Although these exhibit $\delta V < -0.25$, the absence of a double-peaked $^{12}$CO profile indicates that the negative asymmetry likely results from tracer-dependent velocity offsets, i.e., small systematic shifts between the centroid velocities of different molecular tracers caused by variations in optical depth and excitation conditions along the line of sight, rather than true infall dynamics. The average spectra of the remaining blue-asymmetric regions are shown in the top-middle right panel, where the $^{13}$CO (in blue) red-peak at around $-0.7$ km s$^{-1}$ is located near the dip between blue ($-1.0$ km s$^{-1}$) and red ($0.2$ km s$^{-1}$) peaks in the $^{12}$CO profile (in black), and the blue peak dominates—consistent with classical infall signatures.

On the other hand, in some red-asymmetric regions, both tracers display single-peaked lines, but a redward shift in $^{12}$CO relative to $^{13}$CO results in $\delta V > 0.25$. These are more likely attributable to tracer-dependent velocity offsets rather than genuine expansion features. The bottom-upper right panel displays the averaged spectra for the remaining red-asymmetric regions. In these cases, the $^{13}$CO (in blue) blue peak at around $-1.0$ km s$^{-1}$ lies near the dip between blue ($-1.7$ km s$^{-1}$) and red ($-0.7$ km s$^{-1}$) peaks in the $^{12}$CO profile, where the red peak dominates, pointing to potential expansion motions.

In the NE tail, there is one region which lacks a self-absorbed $^{12}$CO profile typically associated with infall, despite a negative $\delta V (<-0.25)$, suggesting the asymmetry is again due to tracer-dependent velocity offsets. Similarly, in another region, the $^{13}$CO peak coincides with the blue component of the $^{12}$CO line rather than the dip, weakening the case for infall. The top-lower right panel shows the averaged profiles for the remaining blue-asymmetric NE tail regions, where the $^{13}$CO red-peak (0.6 km s$^{-1}$) resides near the dip between the blue (0.0 km s$^{-1}$) and the red (1.3 km s$^{-1}$) in the $^{12}$CO profile—supporting infall. The bottom-lower right panel presents the averaged spectra of red-asymmetric regions in the NE tail, showing characteristic expansion signatures where the $^{13}$CO blue-peak (0.3 km s$^{-1}$) lies in the dip between red (1.3 km s$^{-1}$) and blue (0.0 km s$^{-1}$) peaks of $^{12}$CO. In both analyses, the red and blue components are defined relative to the systemic velocities of the NW and NE tails, as derived from the $^{13}$CO emission and discussed in Section~\ref{subsec:Examining Moment Maps and Spectral Features}.

To statistically evaluate the kinematic asymmetry across the regions, \citet{mardones1997search} introduced the blue-excess parameter:
\begin{equation}
    E = \frac{N_{-} - N_{+}}{N},
\end{equation}
where $N_{-}$ and $N_{+}$ denote the number of blue- and red-asymmetric profiles, respectively, and $N$ is the total number of regions. For the NW tail, we find $E = -0.23$, indicating a predominance of red-asymmetric profiles and thus suggesting large-scale expansion, possibly driven by external feedback processes rather than gravitational infall.

In the NE tail, region 14 presents two $^{12}$CO components of nearly equal intensity, making its asymmetry ambiguous; hence, it is excluded from the blue-excess calculation. The remaining regions yield a blue-excess value of $E = -0.22$, also indicating a statistically significant excess of red-asymmetric profiles. This further supports the scenario of expansion dominating the kinematic behavior in the northeastern part of the cloud.

\section{Discussion}
 
The primary aim of this study has been to investigate the star formation potential of the tail regions of SFO 38 and to assess how radiatively driven implosion regulates star formation differently in the head and tails of the cloud. In particular, we seek to determine whether the material channeled from the dense southern head into the tails can accumulate sufficiently to trigger new star formation, or whether it is instead redistributed and dispersed, thereby suppressing gravitational collapse and inhibiting star formation.

Our analysis has yielded several key insights. First, the virial analysis presented in Section~\ref{section: mass and stability} and Section~\ref{section: mass of tail} reveals a striking dynamical contrast: while the southern head of SFO 38 is gravitationally bound, both the NW and NE tails are gravitationally unbound. This distinction underscores a fundamental structural and evolutionary difference between the dense head and the more diffuse tails. Second, the $^{12}$CO line profiles across the tails exhibit systematic red asymmetry and a negative blue-excess parameter (see Section~\ref{section: Spectral Profile-Based Gas Kinematics Study of the Northwestern and Northeastern Tails of SFO 38}), signatures that are consistent with expansion-dominated kinematics rather than infall. Such red-skewed profiles are naturally explained in terms of photoevaporative flows driven by the intense UV radiation from the nearby O-type star HD 206267, as anticipated by the RDI framework \citep{lefloch1994cometary}. The associated pressure gradients accelerate the gas outward, imprinting red-asymmetric profiles in optically thick tracers even in the absence of internal protostellar activity. Finally, we find no observational evidence for ongoing star formation in the tails, such as young stellar objects or molecular outflows. Collectively, these results suggest that, although the southern head is actively forming stars, both tails remain quiescent with respect to star formation presently.

Independent studies have proposed that gas is being redistributed from the dense southern head into the NW tail \citep{patel1995large, okada2024bright}. Recent polarimetric observations of the magnetic field by \citet{sugitani2025structure} further suggest that anisotropic dispersal is facilitated along ordered field lines, thereby enhancing the efficiency of head-to-tail mass transport. 

To quantify this process, we employ the cylindrical filamentary model introduced by \citet{kirk2013filamentary}. Within this framework, the mass accretion (or flow) rate can be estimated as the product of the longitudinal velocity component and the filament’s mass per unit length. The effective radius sets the filament’s cross-sectional area, while the total mass and length determine the mean mass density. For an inclined geometry, the observed length and velocity relate to their intrinsic values through the inclination angle, $\alpha$, such that the observed length is given by $L_{\rm obs} = L \cos \alpha$ and the observed line-of-sight velocity by $V_{\parallel,{\rm obs}} = V_{\parallel} \sin \alpha$. Combining these relations, the directed mass transfer rate can be written as: 

\begin{equation}
    \dot{M}_{\parallel} = \frac{\nabla V_{\parallel, obs} \, M}{\tan(\alpha)},
    \label{eq: mass infall rate}
\end{equation}
where $\nabla V_{\parallel, obs}$ is the observed velocity gradient (measured from a linear fit to the variation in $V_{\rm LSR}$ as $1.05$~km~s$^{-1}$~pc$^{-1}$), and $\alpha$ is the inclination angle of the NW tail relative to the plane of the sky, here adopted as $45^\circ$ \citep{rawat2024giant}. Substituting these values yields a directed mass transfer rate from the southern head to the NW tail of $3.6\times10^{-4}$~M$_\odot$~yr$^{-1}$. This rate is substantial and comparable to mass flow rates inferred in other filamentary environments—for example, Mon R2 \citep[$10^{-4}$ M$_\odot$ yr$^{-1}$;][]{trevino2019dynamics}, Serpens \citep[1--3 $\times$ 10$^{-4}$ M$_\odot$ yr$^{-1}$;][]{kirk2013filamentary}, Orion \citep[0.6 $\times$ 10$^{-4}$ M$_\odot$ yr$^{-1}$;][]{hacar2017gravitational}, and Perseus \citep[0.1--0.4 $\times$ 10$^{-4}$ M$_\odot$ yr$^{-1}$;][]{hacar2017fibers}. 

\begin{figure*}
\begin{center}
\resizebox{12.0cm}{8.0cm}{\includegraphics{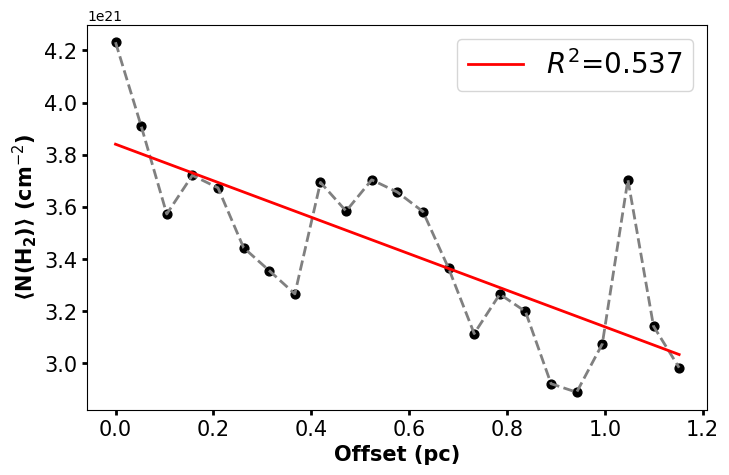}}
\caption{Spatial variation of the mean H$_2$ column density along the NW tail of SFO 38. The mean column density is evaluated at regular intervals of 0.05 pc along this tail, tracing the systematic change in molecular gas distribution. The red line represents a linear regression fit to the data points, with a coefficient of determination R$^{2}$$\approx 0.54$, indicating a moderately strong correlation.}\label{Fig: cd variation}
\end{center}
\end{figure*}

Two possible evolutionary scenarios emerge for the NW tail. It may become supercritical through the accumulation of mass channeled from the dense southern head, or the incoming material could instead be redistributed along the tail, primarily by an increase in its length rather than a significant enhancement in line mass. Examination of the observed hydrogen column density along the NW tail (Figure~\ref{Fig: cd variation}) reveals a declining trend from the base near the head toward the distal end. The red line in the figure, representing a linear fit, aligns with approximately 54\% of the data points, indicating that while mass is indeed flowing into the tail from the head, it is predominantly redistributed longitudinally rather than accumulating sufficiently to render the filament supercritical.

Furthermore, given that the NW tail’s line mass remains below the critical threshold for gravitational instability, we estimate that the minimum timescale required for it to become supercritical is roughly $0.42$~Myr. By contrast, the free-fall timescale of the gravitationally bound southern head, calculated as $t_{\rm ff} = \sqrt{\frac{3\pi}{32 G \rho}}$, where G is the gravitational constant and $\rho$ is the average density, is $t_{\rm ff} \approx 0.37$~Myr. Simultaneously, both the head and the NE tail are subjected to intense ionizing radiation from HD 206267 \citep{choudhury2010triggered}, as discussed in Section~\ref{subsec:Examining Moment Maps and Spectral Features}. Considering solely photoevaporative dispersal, the characteristic evaporation timescale of the head can be estimated following \cite{reipurth1983star, patel1995large}:
\begin{equation}
    t_{\rm ev} = 5000\, n(H_{2}) \, r^{3/2},
    \label{eq: evaporation time}
\end{equation}
where $r$ is the effective radius of the head in parsecs. Our calculations yield $t_{\rm ev} \sim 11$~Myr, consistent with the 10~Myr estimate of \citet{patel1995large}. This comparison highlights that the head’s free-fall timescale is significantly shorter than its dispersal timescale, confirming that it can continue forming stars efficiently prior to complete photoevaporation. Conversely, the NW tail’s supercritical timescale exceeds the head’s free-fall time, implying that it is unlikely to accumulate sufficient mass for gravitational collapse before dispersal dominates.

From these considerations, we conclude that the NW tail, currently gravitationally unbound, is unlikely to evolve into a supercritical state, with its material largely redistributed along its length despite ongoing accretion from the head. The NE tail, which shows no signs of mass inflow, is similarly incapable of reaching supercritical conditions. This suggests that, although mass is channeled from the southern head, the intrinsically unbound nature of the tails inhibits the accumulation of sufficient material to trigger gravitational collapse. Comparable behaviour has been observed in other Galactic cometary globules: CG 1, located in the Gum Nebula region, is more evolved and has already undergone multiple generations of star formation. Radiatively driven implosion in CG 1 has had sufficient time to compress material not only in the head but also downstream into the tail, allowing dense clumps along the tail to fragment and form low-mass stars (IRS 2 and IRS 3; \citealt{makela2012star}). In contrast, CG 2, also in the Gum Nebula, is less evolved, with only a single YSO in the head and a low-density tail that has not yet been significantly affected by RDI. Consequently, its tail remains below the critical density for gravitational collapse. These contrasting cases reinforce the view that tail star formation depends on both the evolutionary stage of the globule and the efficiency of RDI in compressing material into self-gravitating structures.

Collectively, these findings indicate that, although the dense southern head of SFO 38 remains primed for ongoing and future star formation, the NW and NE tails are unlikely to develop into star-forming sites. This behaviour underscores the dual role of RDI in shaping the evolution of the BRC: it efficiently triggers collapse and star formation within the gravitationally bound head, while simultaneously suppressing star formation in the tails by dispersing material and promoting longitudinal redistribution of gas along their lengths.

\section{Summary}
\label{section: summary}

In summary, our study demonstrates that SFO 38 exhibits a clear morphological and dynamical dichotomy between its dense southern head and the more diffuse NW and NE tails. The southern head, particularly its central clump, is gravitationally bound with virial parameters below 2, and shows clear signs of ongoing star formation, likely triggered and sustained by radiatively driven implosion. In contrast, the NW and NE tails are presently gravitationally unbound. The NE tail, receiving no mass from the head, remains quiescent with no evidence for current or imminent star formation. The NW tail, although currently inactive, is subject to mass transfer from the southern head. However, our analysis of the hydrogen column density indicates that this inflowing material is predominantly redistributed along the tail's length rather than increasing the line mass sufficiently to reach a supercritical state. Consequently, the NW tail may not achieve gravitational collapse within relevant timescales, and star formation there is unlikely in the near future. Collectively, these findings reveal the dual and contrasting role of RDI in SFO 38: it promotes star formation within the dense, gravitationally bound head while simultaneously inhibiting collapse in the tails through dispersal and longitudinal redistribution of gas. This underscores the complex and multifaceted influence of radiative feedback in shaping the star formation landscape of cometary globules.

\begin{acknowledgments}
We express our gratitude for the JCMT HARP data utilized in this study, obtained under project code M08BU15. The JCMT is operated by the East Asian Observatory in collaboration with its esteemed partners: the National Astronomical Observatory of China, the National Astronomical Observatory of Japan, the Korea Astronomy and Space Science Institute (KASI), the Academia Sinica Institute of Astronomy and Astrophysics of Taiwan, and the Science and Technology Facilities Council of the United Kingdom.

Furthermore, we acknowledge the publicly available WISE 12 and 4 $\mu$m emission data, accessed through NASA's SkyView Virtual Observatory (\url{https://skyview.gsfc.nasa.gov}), which contributed significantly to this research.

This study was supported by the Indian Institute of Astrophysics (IIA) under the Department of Science and Technology (DST), Government of India. E.J.C. acknowledges support from the Basic Science Research Program, funded by the National Research Foundation of Korea (NRF) through the Ministry of Education (grant number NRF-2022R1I1A1A01053862). C.W.L. is supported by the Basic Science Research Program, also funded by the NRF under the Ministry of Education, Science and Technology (NRF-2019R1A2C1010851), as well as by the Korea Astronomy and Space Science Institute grant, supported by the Korean government (MSIT; project No. 2024-1-841-00).
\end{acknowledgments}

\software{astropy \citep{2013A&A...558A..33A,2018AJ....156..123A},  
          SciPy \citep{virtanen2020scipy}, 
          NumPy \citep{harris2020array}.}

\bibliography{sample701}{}
\bibliographystyle{aasjournalv7}



\end{document}